\documentclass[letterpaper, 10 pt, conference]{ieeeconf}  
\usepackage{graphicx}
\usepackage{booktabs}

\IEEEoverridecommandlockouts                              

\title{\LARGE \bf
QLPro: Automated Code Vulnerability Discovery via LLM and Static Code Analysis Integration
}

\author{Junze Hu$^{1,2}$ , Xiangyu Jin$^{1,2}$ , Yizhe Zeng$^{1,2}$ , Yuling Liu$^{1,2}$ , Yunpeng Li$^{1,2}$ , \\ Dan Du$^{1,2}$ , Kaiyu Xie$^{1,2}$ , Hongsong Zhu$^{1,2,*}$ 
\thanks{$^{*}$Corresponding author}%
\thanks{$^{1}$Institute of Information Engineering, Chinese Academy of Sciences, Beijing, China
        {\tt\small \{hujunze, zengyizhe\}@iie.ac.cn}}%
\thanks{$^{2}$School of Cyber Security, University of Chinese Academy of Sciences, Beiiing, China
        {\tt\small hujunze24@mails.ucas.ac.cn}}%
}

\begin{document}

\maketitle
\thispagestyle{empty}
\pagestyle{empty}

\begin{abstract}

Code auditing, a method where security researchers review source code to identify vulnerabilities, has become increasingly impractical for large-scale open-source projects. While Large Language Models (LLMs) demonstrate impressive code generation capabilities, they are constrained by limitations in context window size, memory capacity, and complex reasoning abilities, making direct vulnerability detection across entire projects infeasible. Static code analysis tools, though effective to a degree, are heavily reliant on their predefined scanning rules. To address these challenges, we present QLPro, a vulnerability detection framework that systematically integrates LLMs with static code analysis tools. QLPro introduces both a triple-voting mechanism and a three-role mechanism to enable fully automated vulnerability detection across entire open-source projects without human intervention. Specifically, QLPro first utilizes static analysis tools to extract all taint specifications from a project, then employs LLMs and the triple-voting mechanism to classify and match these taint specifications, thereby enhancing both the accuracy and appropriateness of taint specification classification. Finally, QLPro leverages LLMs with the three-role mechanism to develop scanning rules based on the matched taint specifications, significantly improving the grammatical correctness of vulnerability scanning rules generated by LLMs. For evaluation, we constructed a new dataset, JavaTest, comprising 10 open-source projects from GitHub with 62 confirmed vulnerabilities. CodeQL, a state-of-the-art static analysis tool, detected only 24 of these vulnerabilities. In contrast, QLPro, using a collaborative approach with two Claude-3.7-thinking, detected 41. Furthermore, QLPro discovered 6 previously unknown vulnerabilities, 2 of which have been confirmed as 0-days, underscoring its ability to identify vulnerabilities beyond the capabilities of existing tools.

\end{abstract}

\section{INTRODUCTION}
Code auditing, while effective for discovering high-severity vulnerabilities through manual source code review, has become increasingly impractical as modern open-source projects grow in scale and complexity. Static code analysis tools, particularly those employing taint analysis, have emerged as crucial solutions to these challenges. Tools like GitHub CodeQL\cite{c1}, Checker Framework\cite{c2}, Snyk Code\cite{c3}, and SonarQube\cite{c4} have gained widespread adoption across various programming languages by enabling automated vulnerability detection through customized rules. The recent researches demonstrates their effectiveness: Predator\cite{c5} identified numerous vulnerabilities in PHP projects; Gobbi et al.\cite{c6} used CodeQL to detect defects in npm packages; and Lipp et al.\cite{c7} evaluated multiple static analysis tools for C/C++ projects.

Despite their widespread adoption, mainstream static code analysis tools predominantly rely on built-in official rule libraries for vulnerability detection. These rules are intentionally designed with generality in mind, resulting in highly generalized approaches that primarily focus on identifying patterned, obvious low-level vulnerabilities. This characteristic makes these tools inadequate for adapting to the specific code context, business logic, and potential unique risk points of individual projects, thereby failing to effectively support deep and precise vulnerability mining within project codebases.

To address the aforementioned lack of specificity in built-in rules, the currently prevalent solution relies on security experts manually writing (or custom developing) specialized scanning rules for specific projects. While this approach can, to some extent, enhance the alignment between rules and project code—thereby uncovering more concealed and threatening vulnerabilities—it faces the following limitations: \textbf{L1:} \textit{Expert Dependency.} Custom rule quality depends directly on the author's expertise, vulnerability pattern knowledge, and familiarity with the target codebase. \textbf{L2:} \textit{Resource Intensity.} Rule design, development, testing and maintenance demand significant professional time and human resources. \textbf{L3:} \textit{Effectiveness Uncertainty.} Despite substantial investment, there remains no guarantee that rules will successfully detect unknown vulnerabilities or provide comprehensive coverage of potential risk points. Consequently, we aim to develop a methodology that can automatically generate appropriate vulnerability scanning rules for individual projects, thereby enabling even non-security professionals to efficiently and accurately identify potential vulnerabilities within projects.

Large Language Models (LLMs) have emerged as a recent technological focal point due to their potential in general artificial intelligence and ability to support multiple downstream tasks. Numerous studies have demonstrated impressive capabilities of these models in code generation. For instance, Chen et al.\cite{c8} developed the Codex model based on GPT-3, which became the core technology powering GitHub Copilot; Fried et al.\cite{c9} demonstrated capabilities in code generation tasks involving long-range contextual dependencies; and Anthropic's Claude 3 model\cite{c10} exhibited excellent performance in zero-shot code generation tasks. Unfortunately, despite these advancements, significant challenges persist when attempting to use LLMs for the automatic generation of vulnerability scanning rules:

\textbf{C1:} \textit{The effectiveness of LLM-generated vulnerability scanning rules is critically dependent on the quality of input static taint specifications.} While some expect LLMs to generate effective vulnerability scanning rules from high-level descriptions alone, practical experience shows this approach is insufficient. Effective rules require precise characterization of taint propagation paths—accurately identifying sources, sinks, and their relationships within specific code contexts. Static taint specifications carry this critical information, and LLMs' rule-generation effectiveness is fundamentally limited by the classification accuracy and matching appropriateness of the taint specifications they receive.

\textbf{C2:} \textit{The one-shot use of LLMs for code generation in niche programming languages leads to limited performance.} The most intuitive approach to code generation involves providing direct prompts to the LLM, such as ``Please write vulnerability scanning rules that can detect SQL injection based on the given taint specification", and then collecting the response. However, in practice, we find that vulnerability scanning rules used by current static analysis tools are predominantly written using tool-specific syntax structures. This prevents LLMs from generating directly compilable code files without human intervention, unlike when writing in mainstream programming languages such as Java, Python, or C/C++. The recommended approach for utilizing LLMs effectively is to decompose the task into distinct components and guide the model incrementally, thereby maximizing its ability to produce vulnerability scanning rules without syntax errors\cite{c11}.

\textbf{QLPro.} In this paper, we present QLPro, a novel framework for fully automated generation of vulnerability scanning rules targeting individual projects (using CodeQL static code analysis tool in our experiments). QLPro can be directly applied to complete standalone open-source projects, supporting automatic extraction of taint specifications with classification, matching, and ultimately generating comprehensive vulnerability scanning files based on the provided taint specification tuples. QLPro enables both seasoned security researchers and junior developers without security experience to efficiently identify potentially vulnerable code within projects. Specifically, we address the aforementioned challenges through the following insights. First, regarding \textbf{C1}, we extract all taint specifications from the project using CodeQL, then leverage large language models with a triple-voting mechanism to accurately classify these specifications into sources, sinks, or none. We then match appropriate sources and sinks to form (source, sink) binary tuples. Second, for \textbf{C2}, we propose a three-role mechanism (Writer, Repair, Execute) to enhance syntactic correctness of LLM-generated QL files. Writer creates files from (source, sink) tuples; Execute attempts compilation; Repair suggests modifications for failed compilations until success or reaching maximum iteration limit. Experimental results from section IV-C and prior research\cite{c12} confirm this approach significantly improves both QL file accuracy and vulnerability detection capabilities.

\textbf{Contributions.} Our contributions are summarized as follows: \textbf{\textit{Approach.}} We present the first integrated framework that combines Large Language Models and static code analysis tools to achieve fully automated vulnerability scanning rule generation. This framework addresses several key challenges that persisted in previous research through its voting mechanism and three-role mechanism, thereby maximizing the potential of LLMs. Our approach makes it possible for non-security professionals to efficiently and accurately identify potential vulnerabilities across entire open-source projects. \textbf{\textit{Dataset.}} To evaluate our approach, we curated JavaTest, a compilable Java project dataset. JavaTest comprises ten widely-used, open-source Java projects, each averaging 10.53K stars on GitHub and collectively containing 62 documented vulnerabilities of various types. The projects within this dataset exhibit significant complexity, with an average of 50K lines of code (LOC) per project, and some projects exceeding 100K LOC. This complexity establishes JavaTest as a challenging benchmark dataset for vulnerability scanning. \textbf{\textit{Result.}} We evaluated QLPro on the JavaTest, conducting both cross-model and within-model comparative experiments using four distinct large language models. The results demonstrate that QLPro, when integrated with two Claude-3.7-Sonnet-Thinking, not only achieves the highest CodeQL syntax correctness rate (90.90\%) but also identifies a significantly greater number of vulnerabilities (41) compared to using the official CodeQL rule library alone (24). Furthermore, QLPro discovered 6 previously unknown vulnerabilities (0-days), providing compelling evidence of the method's superior performance.

\section{Preliminary}
Generally, leveraging large language models (LLMs) for the creation of custom rules for static analysis tools involves three primary steps: \textit{1) Extraction and Classification of Sources and Sinks}, \textit{2) Source-Sink Pairing} and \textit{3) Query Rule Generation}.

\textit{1) Extraction and Classification of Sources and Sinks:} Sources and sinks are fundamental concepts in taint tracking. A source represents the entry point of untrusted data into a program, often a point where an attacker could potentially control inputs.  In CodeQL, a source serves as the starting point for data flow analysis. Conversely, a sink represents a potentially dangerous operation or location in the code where untrusted data could trigger a vulnerability. In CodeQL, a sink serves as the endpoint for data flow analysis. 

\textit{2) Source-Sink Pairing:} Valid sources and sinks typically occur in pairs. Only a correct (source, sink) pair can trigger a specific vulnerability. For instance, in the context of SQL injection, avalid pair might be (getParameter(), executeQuery()) where getParameter() retrieves untrusted input and executeQuery() executes a potentially vulnerable SQL query. 

\textit{3) Query Rule Generation:} The objective of this step is the automated generation of query rules by the LLM, given a specific (source, sink) pair.  That is, the input is a (source, sink) tuple, and the desired output is a complete, syntactically correct QL code file.

\begin{figure*}[htbp]
  \centering
  \includegraphics[scale=0.2]{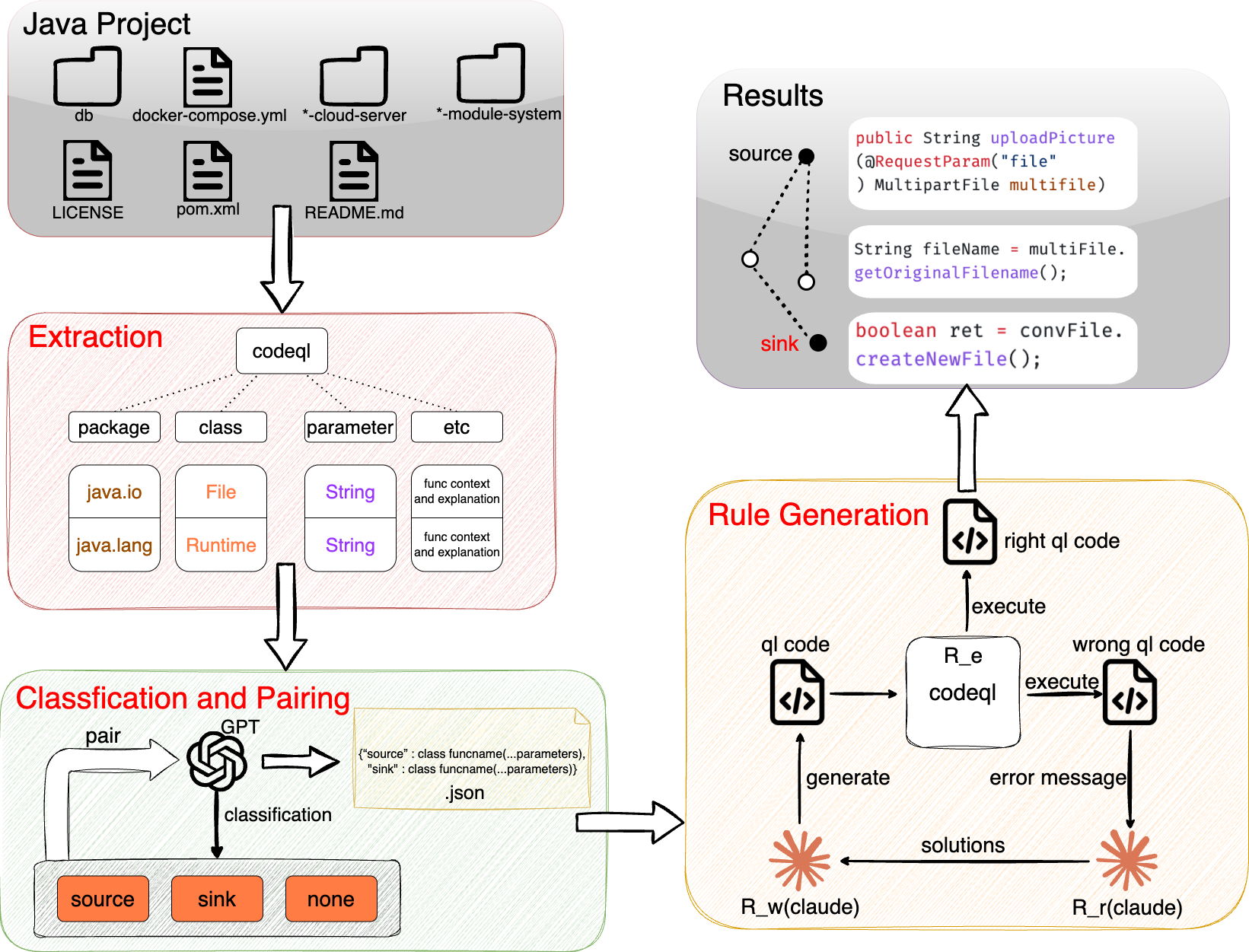}
  \caption{Workflow of QLPro. Including three modules: Extraction, Classification and Matching, Rule Generation. We use R\_w to represent the write, R\_e to represent the execute, and R\_r to represent the repair for readability in the following sections.}
  \label{overview}
\end{figure*}

\section{QLPro Framework}
In this section, we introduce the design of QLPro, a framework that utilizes large language models to automatically generate targeted CodeQL query rules for individual open-source projects. As illustrated in \textbf{Fig.1}, the framework consists of three main components: Extraction, which acquires all APIs from the project through CodeQL; Classification and Matching, which employs large language models and a triple-voting mechanism to categorize APIs into sources and sinks, then forms appropriate (source, sink) binary tuples; and Rule Generation, which leverages large language models and a three-role mechanism to automatically generate QL query rule files for the given (source, sink) tuples.

\begin{figure*}[htbp]
  \centering
  \includegraphics[scale=0.19]{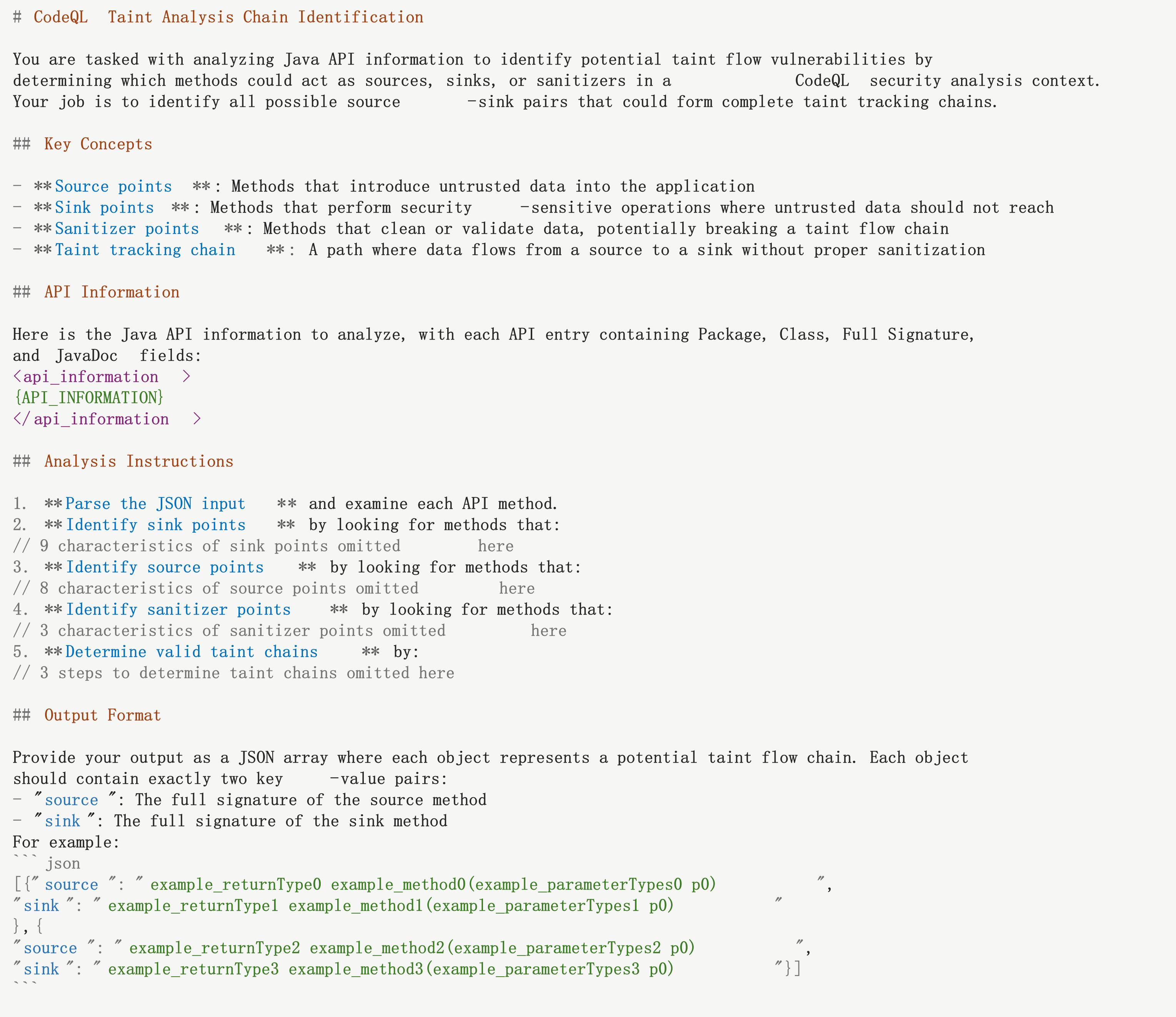}
  \caption{An example prompt for classifying taint specifications, designed to categorize them into sources and sinks.}
  \label{classification}
\end{figure*}

\subsection{Extraction}

The purpose of this step is to extract potentially dangerous function calls from a Java open-source project. We first use CodeQL to extract all APIs from the project, including package names, parameter lists, return types, annotation information, and other metadata. We then remove function calls that are unlikely to pose security risks. Simultaneously, we eliminate duplicate function calls, as these typically appear in different files or locations, and our focus is solely on the function characteristics and relevant code fragments rather than their invocation locations. Finally, we construct this information into JSON format, as this representation combined with partial function bodies can effectively characterize the distinctive features of a function.

\subsection{Classification and Pairing}

Given the exceptional potential of Large Language Models (LLMs) in general artificial intelligence, this research employs refined prompt engineering methods combined with a triple-voting mechanism to achieve efficient classification and matching of sources and sinks. Initially, we designed a structured prompt template as illustrated in \textbf{Fig.2}, guiding the model to perform semantic analysis and classification of JSON data containing API information. In our prompt construction, we begin by establishing the primary objective for the LLM: to classify and pair source and sink points from the provided API dataset. We then precisely define five critical concepts essential to this analysis: source points, sink points, sanitizer points, and taint tracking chains. Following these definitions, we input the data to be processed, designated as API\_INFORMATION. The prompt subsequently delineates a systematic analytical framework comprising six discrete steps: 1) Parse the input JSON data structure and methodically examine each API method. 2) Apply 9 distinct characteristics to identify potential sink points within the codebase. 3) Employ 8 defined heuristics to locate potential source points. 4) Utilize 3 specific criteria to recognize sanitizer points that may interrupt taint propagation. 5) Execute a three-phase algorithm to determine valid taint propagation paths between identified sources and sinks. 6) Finally, we impose a rigorous output schema specification, requiring results to be formatted as binary tuples, each representing a validated source-sink pair with potential security implications.

However, considering the limited input capacity of large models, which prevents processing all JSON data simultaneously, we divided the data into groups to ensure each group's length remains within the model's input constraints. This approach introduced a new challenge: the model exhibited significant context interference during processing, where the classification result of an individual API would be semantically influenced by other APIs in the same group, reducing classification precision. To address this challenge, we proposed and implemented a triple-voting mechanism: each API is placed in three distinct contextual environments for independent evaluation, with the final classification determined by majority voting principle, effectively mitigating the impact of contextual bias on classification accuracy. Subsequently, we needed to match the classified sources and sinks. Since not all paths from sources to sinks necessarily exist, we employed large language models to match sources and sinks to construct reasonable (source, sink) binary tuples, aiming to identify function calls most likely to form vulnerability exploitation chains.

\subsection{Rule Generation}

Previous researchers\cite{c12} used predefined templates to write CodeQL statements, but this approach could only address very simple issues and proved inadequate when confronting complex code logic. To overcome this limitation, we explored leveraging large language models to automatically generate CodeQL rules. Furthermore, due to the unique syntax of CodeQL, LLMs cannot achieve near-perfect syntactic accuracy (as they often can with mainstream languages like Java and Python) without intervention. Therefore, we proposed and implemented a three-role mechanism to enhance the syntactic accuracy of generated CodeQL files. Specifically, the three-role mechanism comprises three roles: R\_w, R\_e, and R\_r. R\_w (writer) is responsible for generating corresponding QL files based on the (source,sink) binary tuples provided by the upstream language model and then passing them to R\_e. Upon receiving the QL files generated by R\_w, R\_e (executor) attempts to compile and execute them. If compilation fails, R\_e forwards the QL file along with error information to R\_r (repairer) for validation. R\_r's primary function is to analyze the error information and propose modification suggestions and solutions for the QL file, which are then sent back to R\_w. After receiving the solutions provided by R\_r, R\_w modifies the code in the QL file accordingly and repeats the previous operations until the QL file can be correctly compiled and executed by R\_e. In this process, we established a maximum modification count (MAX); if a QL file still cannot be correctly compiled after MAX modifications, it is considered invalid.

\section{Evaluation}
In this section, we evaluate QLPro by answering the following research questions:

\begin{itemize}

\item RQ1: What is the syntactic correctness rate of CodeQL rules generated by QLPro?
\item RQ2: Compared to the official CodeQL rule repository, can the rules generated by QLPro detect a greater number of known vulnerabilities within the same target project?
\item RQ3: Can QLPro identify novel, previously undocumented vulnerabilities?

\end{itemize}

\subsection{Dataset}

To evaluate the effectiveness of our proposed approach, we required a dataset comprising several complete Java projects. Each project needed to be fully compilable by CodeQL, a fundamental prerequisite for applying static analysis techniques. Furthermore, the projects should originate from real-world applications, as these typically exhibit greater complexity and, consequently, present more significant challenges for analysis. Finally, the vulnerabilities within the projects must be verifiable, allowing for a reliable assessment of the detection tool's performance.

To fulfill these criteria, we assembled the JavaTest dataset, consisting of ten popular, open-source Java projects sourced from GitHub. These ten projects collectively contain 62 confirmed vulnerabilities. The average LOC exceed 50K with the largest project surpassing 100K LOC, indicating a substantial level of complexity and, therefore, posing a considerable challenge for vulnerability detection.

\subsection{Model selection}

For our experiments and evaluation, we selected two closed-source LLMs: Claude-3.7 (version: Claude-3.7-sonnet-2025-0219) and Claude-3.7-thinking (version: Claude-3-7-sonnet-20250219-thinking )\cite{c10}. To provide further comparative analysis, we also included an additional closed-source LLMs, which we refer to as GPT-4o-mini\cite{c13} for the purposes of this study.

\subsection{Results}
\textbf{1) RQ1: Syntactic Correctness Rate of Rules Generated by QLPro}

Due to the unique syntax of CodeQL, LLMs cannot achieve near-perfect syntactic correctness in code generation without specific adaptations, unlike their performance with mainstream languages like Java or Python. Therefore, it is crucial to assess the syntactic correctness rate of QL files generated by LLMs after applying our proposed methodology, using successful compilation as the benchmark for correctness.

We completed the pairing of (source, sink) elements within the dataset, resulting in 1924 QL files, namely 1924 unique (source, sink) pairs. Subsequently, we conducted the evaluation. \textbf{Table 1} and \textbf{Fig.3} presents the performance of different LLMs in generating syntactically correct QL files. Specifically, when using a single LLM with our carefully crafted prompts, Claude-3.7-sonnet-thinking achieved the highest syntactic correctness rate at 70.01\%. Other LLMs demonstrated comparatively lower correctness rates. Notably, when employing a collaborative approach with two instances of Claude-3.7-sonnet-thinking, the correctness rate significantly improved to 90.90\%. This demonstrates that, given (source, sink) pairs, our method can effectively enable LLMs to generate QL files with a high degree of syntactic correctness.

\begin{table}[ht]
\caption{Syntactic Correctness Rate of Rules Generated by Different Models}
\label{table_example}
\begin{center}
\resizebox{\linewidth}{!}
{
\begin{tabular}{ccc}
\toprule
\textbf{Model} &  \textbf{Correct Number} & \textbf{Correct Rate}\\ 
\midrule
Claude-3.7 & 1347 & 70.01\%\\

Claude-3.7-thinking  & 1452 & 75.47\%\\

Claude-3.7-thinking + Claude-3.7-thinking  & \textbf{1749} & \textbf{90.90}\%\\

Claude-3.7 + GPT-4o-mini  & 1522 & 79.10\%\\
\bottomrule
\end{tabular}
}
\end{center}
\end{table}

\begin{figure}[htbp]
  \centering
  \includegraphics[width=0.49\textwidth]{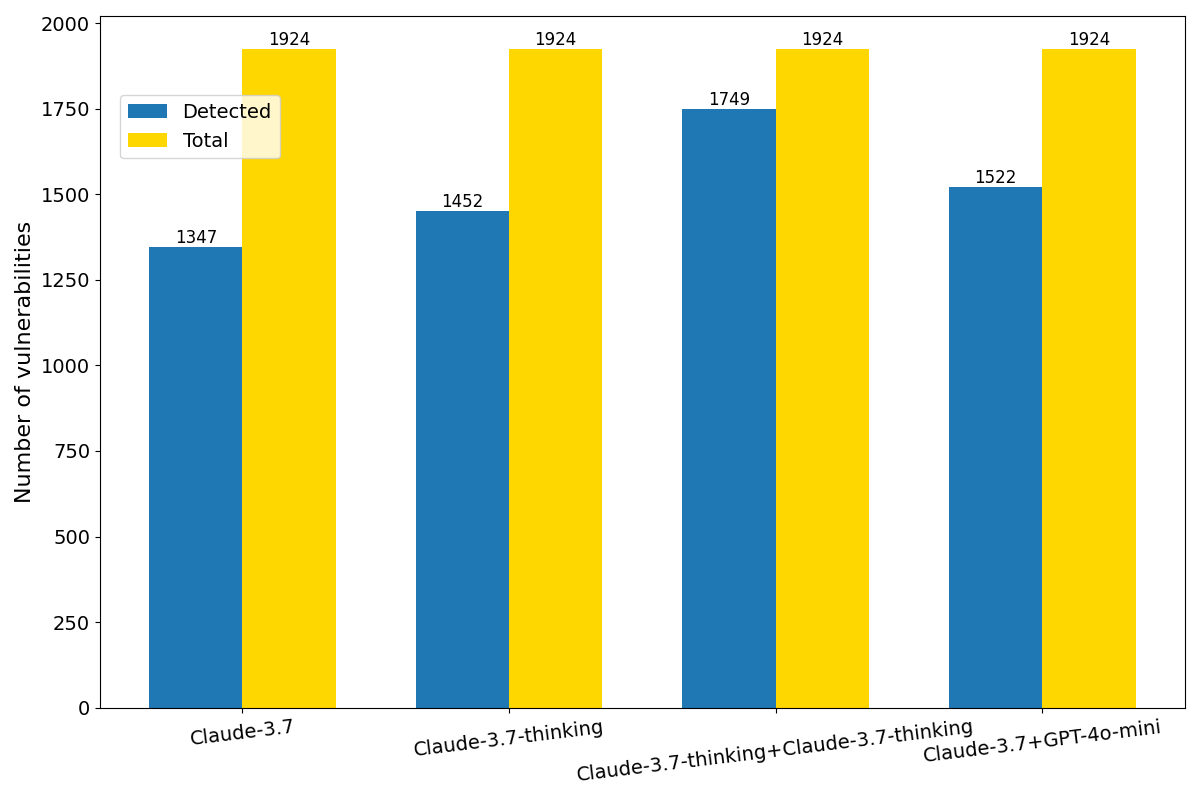}
  \caption{Number of Correct Files Generated by Different Models}
  \label{correct files}
\end{figure}

\textbf{2) RQ2: Effectiveness of QLPro in Detecting Existing Vulnerabilities}

\textbf{Table 2} and \textbf{Fig.4} highlights the comparative performance of rules generated by QLPro versus the official rule repository on the JavaTest benchmark. Specifically, when QLPro was used in conjunction with two Claude-3.7-sonnet-thinking, a total of 41 vulnerabilities were identified. In contrast, the official rule repository identified 24 vulnerabilities. Our analysis suggests that the broader, more generalized definitions of(source, sink)pairs in the official rules led to the omission of several vulnerabilities. These results indicate that our method surpasses the official rule repository in detecting existing vulnerabilities, reducing the likelihood of overlooking potential security issues.

\begin{table}[ht]
\caption{The Effectiveness in Detecting Existing Vulnerabilities by Different Models}
\label{table_example}
\begin{center}
\resizebox{\linewidth}{!}
{
\begin{tabular}{ccc}
\toprule
\textbf{Model} & \textbf{Detected} & \textbf{Detection Rate}\\ 
\midrule
Claude-3.7 & 29 & 46.80\%\\

Claude-3.7-thinking & 29 & 46.80\%\\

Claude-3.7-thinking + Claude-3.7-thinking & \textbf{41} & \textbf{66.10}\%\\

Claude-3.7 + GPT-4o-mini & 31 & 50.00\%\\

CodeQL & 24 & 38.70\%\\
\bottomrule
\end{tabular}
}
\end{center}
\end{table}

\begin{figure}[t]
  \centering
  \includegraphics[width=0.49\textwidth]{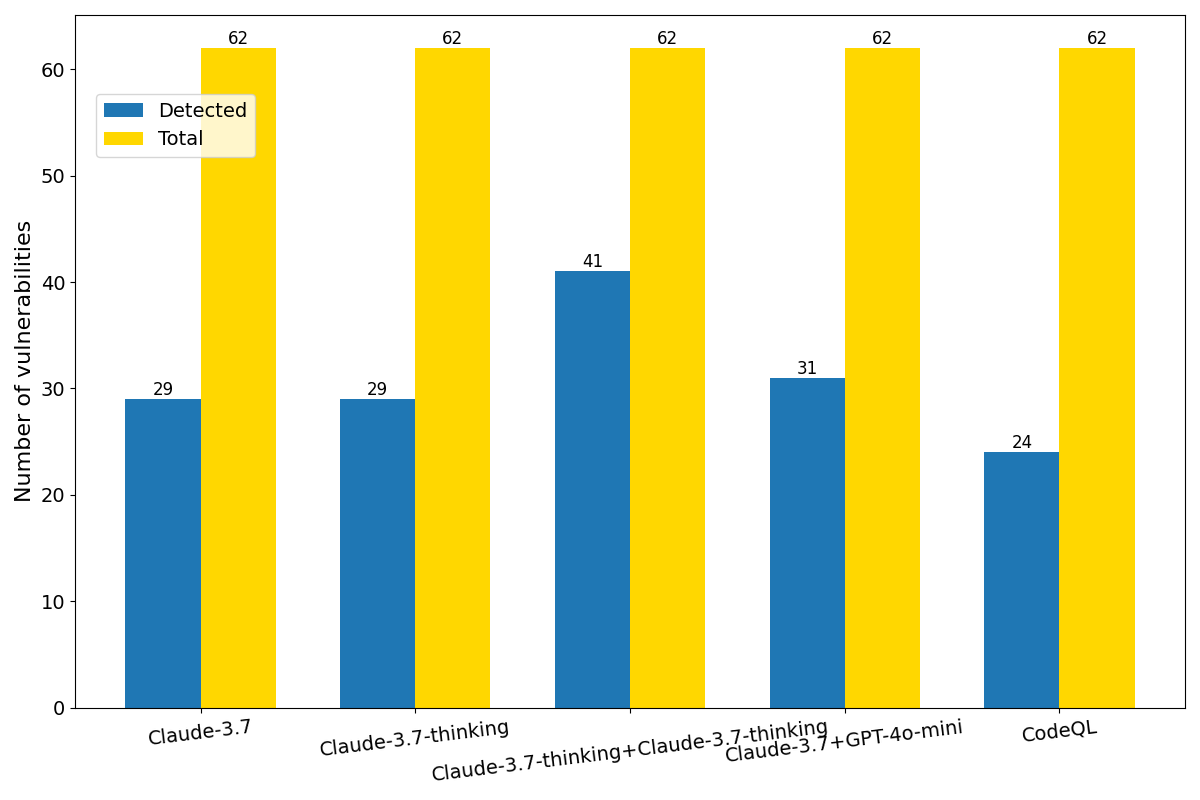}
  \caption{Number of Vulnerabilities Detected by Different Models}
  \label{vulnerabilities}
\end{figure}

\textbf{3) RQ3: Capability of QLPro in Detecting Unknown (0-Day) Vulnerabilities}

We applied both QLPro (integrated with two Claude-3.7-sonnet-thinking) and the official CodeQL rule repository to our dataset. Our analysis revealed that QLPro identified 6 previously unknown vulnerabilities (0-days). We reported these 6 vulnerabilities to the developers and 2 of these have been officially confirmed as 0-day vulnerabilities, and the remaining four are pending confirmation. These vulnerabilities encompass various types, including SQL injection, arbitrary file upload, and arbitrary file read. \textbf{Fig.5} illustrates an example of arbitrary file upload. Through a comparison with the vulnerabilities reported by the official rules, we confirmed that these six vulnerabilities were not detected by the official rule set. This demonstrates the capability of QLPro to detect unknown vulnerabilities.

\begin{figure}[t]
  \centering
  \includegraphics[width=0.49\textwidth]{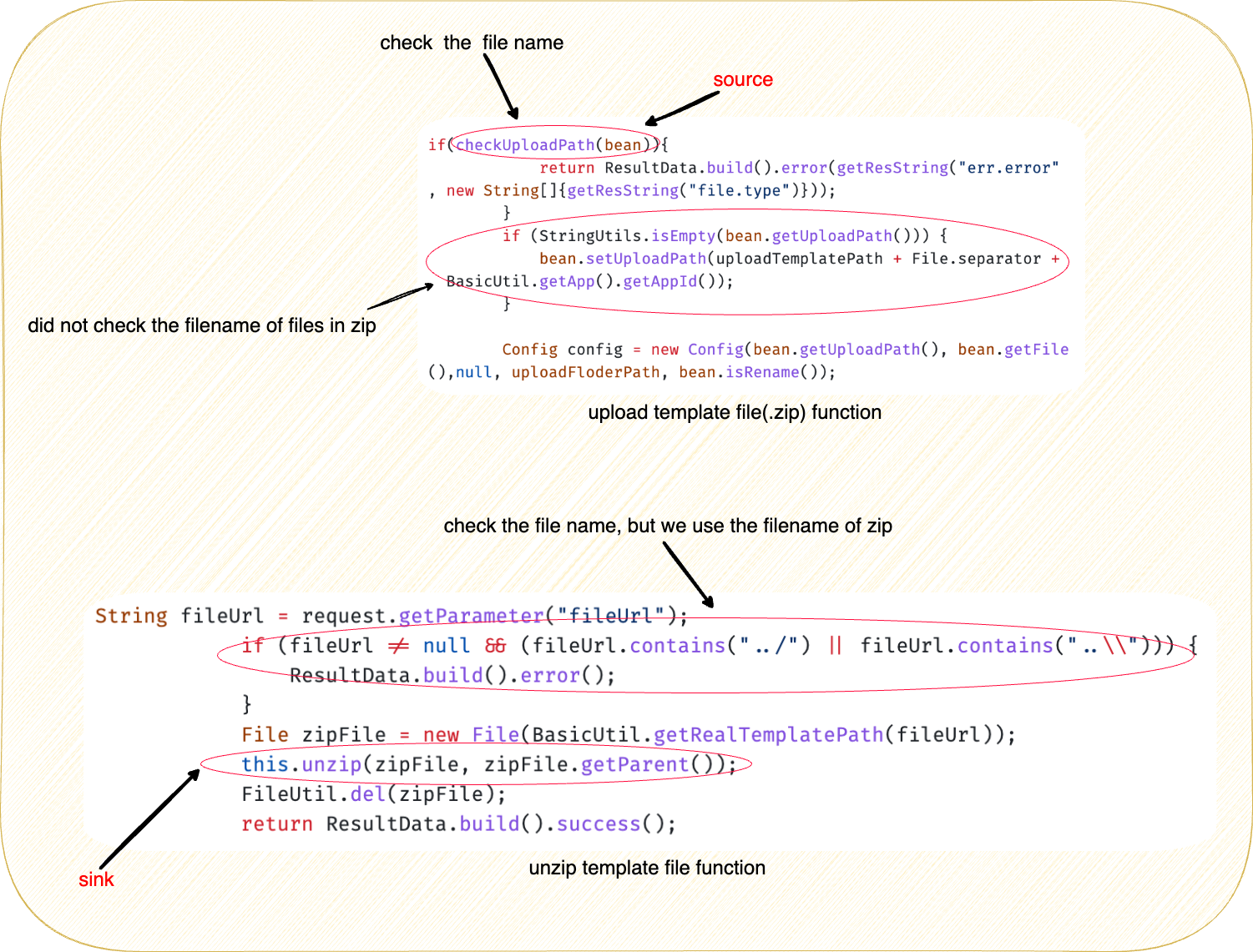}
  \caption{QLPro identified a previously unreported 0-day vulnerability classified as arbitrary file upload. The vulnerability is exploitable through a crafted filename incorporating directory traversal characters (such as ../../1.jsp).  By compressing a file with this manipulated filename into an archive and uploading it, an attacker can achieve path traversal, leading to the unauthorized writing of files to arbitrary locations on the target system}
  \label{file upload}
\end{figure}

\section{Conclusion}

In this paper, we systematically explored how to achieve fully automated code vulnerability detection by integrating Large Language Models with static code analysis tools. To maximize the potential of LLMs, we designed a triple-voting mechanism and a three-role mechanism. Our evaluation of QLPro demonstrates that it successfully detected 41 out of 62 vulnerabilities in the test dataset (66.1\%), significantly outperforming the official CodeQL tool's 24 out of 62 vulnerabilities (38.7\%). Additionally, QLPro discovered 6 previously unknown vulnerabilities (0days). QLPro makes it possible for non-security professionals to efficiently and rapidly identify vulnerabilities in open-source projects, democratizing security analysis capabilities.

\addtolength{\textheight}{-12cm}   




\section*{ACKNOWLEDGMENT}
This study is funded by National Key Research and Development Program of China (No.2023YFC3306305).


\end{document}